\documentclass{appolb}
\usepackage{graphicx}

\usepackage{dcolumn}% Align table columns on decimal point
\usepackage{bm}% bold math
\usepackage[chatter]{rotating}

\def\pp{\mbox{$p$-$p$} }
\def\auau{\mbox{Au-Au} }

\def\aa{\mbox{$A$-$A$} }
\def\nn{\mbox{$N$-$N$} }

\def\pt{$p_t$ }

\def\bea{\begin{eqnarray}}
\def\eea{\end{eqnarray}}

\begin{document}
% \eqsec  % uncomment this line to get equations numbered by (sec.num)
\title{Finding, and not finding,
``higher harmonic flows''%
\thanks{Presented at ISMD 2012, Kielce, Poland, 16-21 September 2012}%
% you can use '\\' to break lines
}
\author{Thomas A.\ Trainor
\address{CENPA 354290, University of Washington, Seattle, WA 98195, USA}
\\
}

\maketitle

\begin{abstract}
Certain analysis methods have emerged recently that claim to reveal ``higher harmonic flows'' in more-central \aa collisions at the RHIC and LHC. But pQCD calculations describe the same structures quantitatively.
\end{abstract}

  
\section{Introduction}

I review methods for obtaining ``higher harmonic  flows'' from correlation data in the context of a two-component model of $p_t$ spectra and correlations and pQCD calculations of spectrum hard components. The main focus is on the mechanism for the same-side (SS) 2D peak in angular correlations.

%persistence of jets is inconvenient for "perfect liquid"

%major effort to convert jet structures to flows

%glasma flux tubes, Fourier series analysis

%detailed study confirms a jet mechanism, pQCD description in all cases

%%%%%%%%%%%%%%%%%%%%%%
\section{Two-component model of \pp collisions}

To understand the structure of high-energy \aa collisions we must first develop an accurate \pp reference. The two-component model of measured 200 GeV \pp spectra and correlations includes a soft component -- projectile nucleon dissociation (proton fragmentation) -- and a hard component -- large-angle-scattered parton fragmentation dominated by 3 GeV gluons. 

\begin{figure}[htb]
\centerline{%
\includegraphics[width=.24\textwidth,height=.24\textwidth]{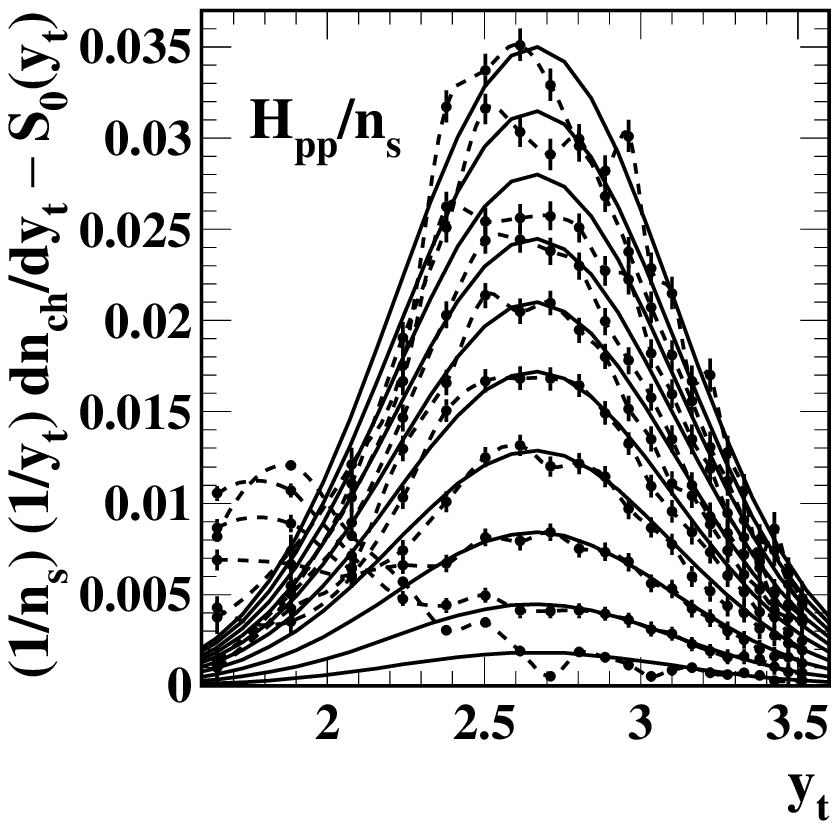}
\includegraphics[width=.24\textwidth,height=.24\textwidth]{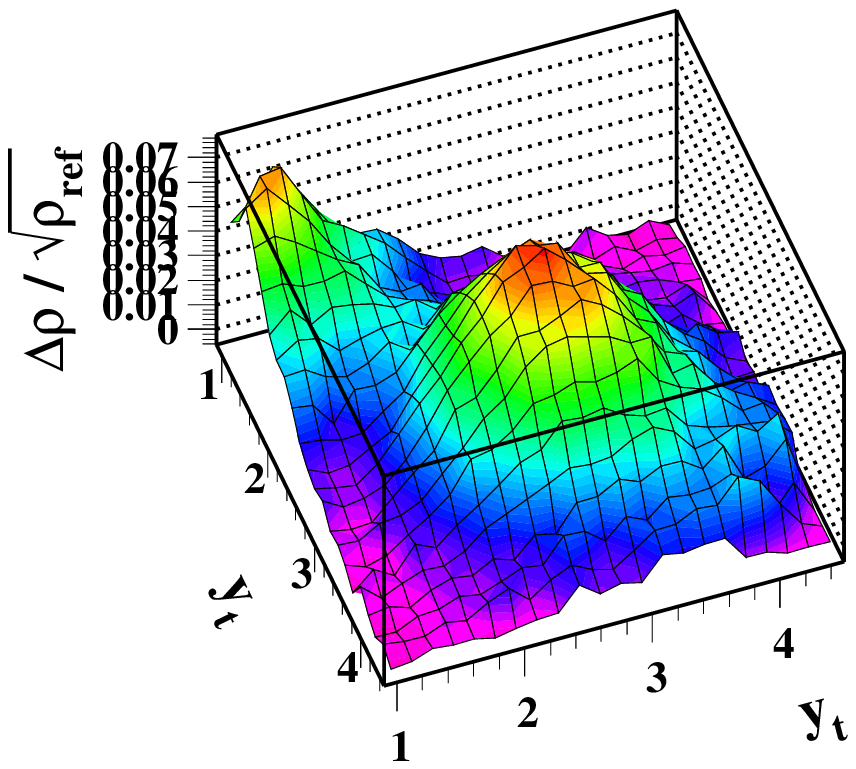}
\includegraphics[width=.24\textwidth,height=.24\textwidth]{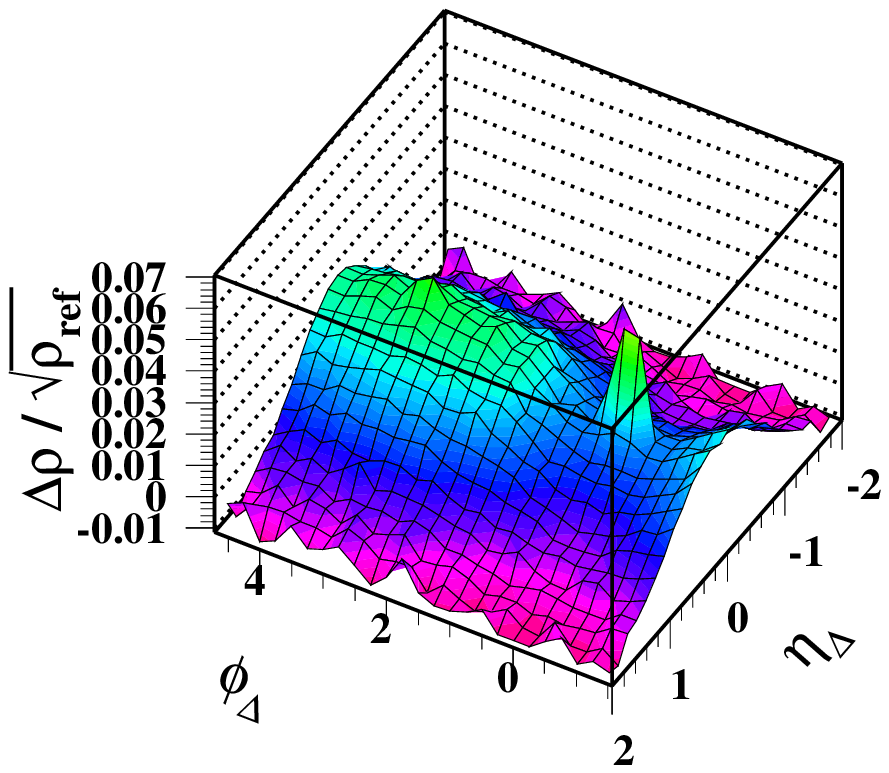}
\includegraphics[width=.24\textwidth,height=.24\textwidth]{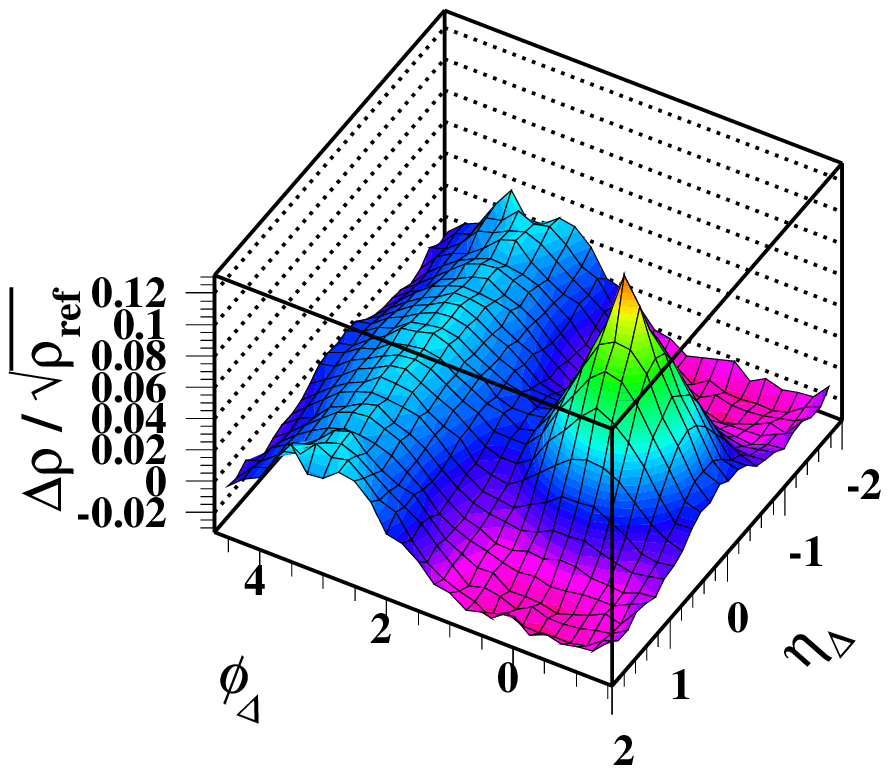}
}
\caption{
First: $y_t$ spectrum hard component vs $n_{ch}$,
Second: Correlations on $(y_t,y_t)$,
Third:  Correlations soft component,
Fourth: Hard component (jet structure).
}
\label{pp2comp}
\end{figure}

Figure~\ref{pp2comp} (first) shows spectrum hard components from 200 GeV \pp collisions on transverse rapidity  $y_t \equiv \ln[(p_t + m_t)/m_\pi]$ which provides balanced visual access to low- and high-\pt structure~\cite{ppprd}. We observe that soft and hard components scale with $n_s$ and $n_s^2$ respectively ($n_s$ is the soft component of multiplicity $n_{ch}$) and are thus easy to distinguish down to 0.5 GeV/c.

Figure~\ref{pp2comp} (second) shows a followup study of two-particle correlations on $(y_t,y_t)$ showing distinct soft and hard components. Figure~\ref{pp2comp} (third, fourth) show corresponding soft and hard components of angular correlations. The data are minimum-bias, with no trigger condition. Structure can be split into same-side (SS) and away-side (AS) on azimuth, with like-sign (LS) and unlike-sign (US) charge combinations. The combined systematics of spectrum and correlation hard components compel a (di)jet interpretation~\cite{porter2,porter3}.

%%%%%%%%%%%
\section{Correlation and spectrum systematics for \auau collisions}

Given a well-defined \pp reference what happens to the two-component model in \aa collisions? Extensive study of \auau spectra and correlations at 62 and 200 GeV establishes centrality and energy dependence~\cite{hardspec,anomalous}. \aa centrality is measured by mean participant path length $\nu = 2 N_{bin}/N_{part}$ obtained from the Glauber model. $\nu$ is used only as a geometry parameter (the \nn cross section is maintained at the 200 GeV value for all cases).

\begin{figure}[htb]
\centerline{%
  \includegraphics[width=.24\textwidth,height=.24\textwidth]{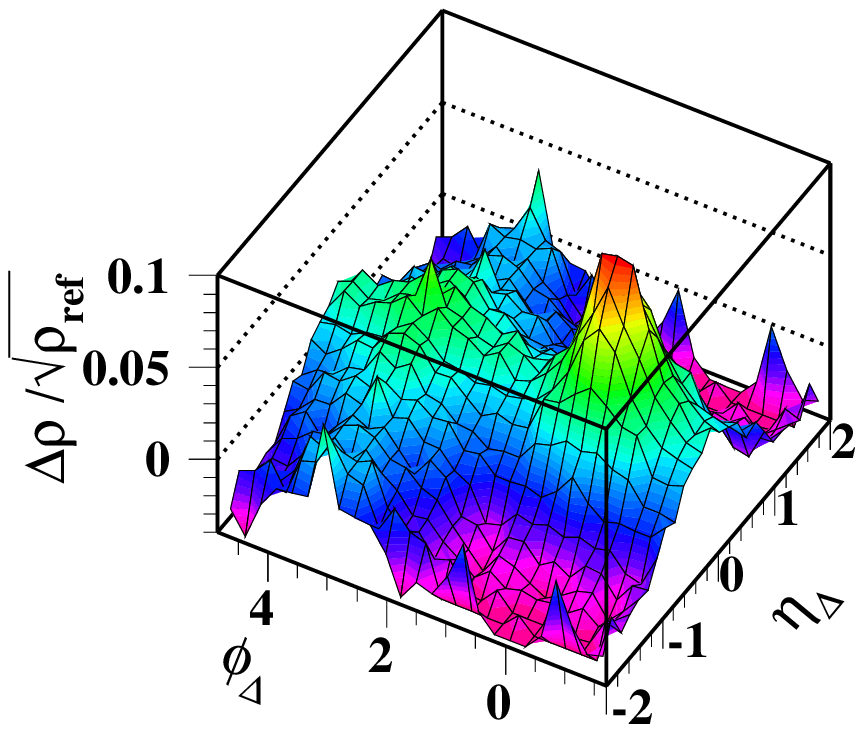}
  \includegraphics[width=.24\textwidth,height=.24\textwidth]{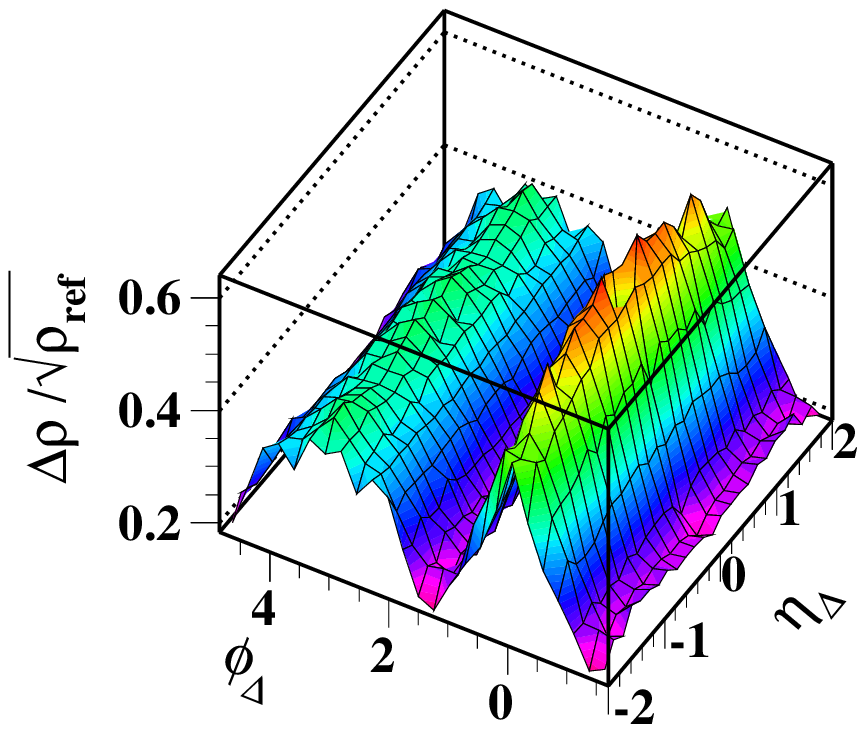}
 \includegraphics[width=.24\textwidth,height=.235\textwidth]{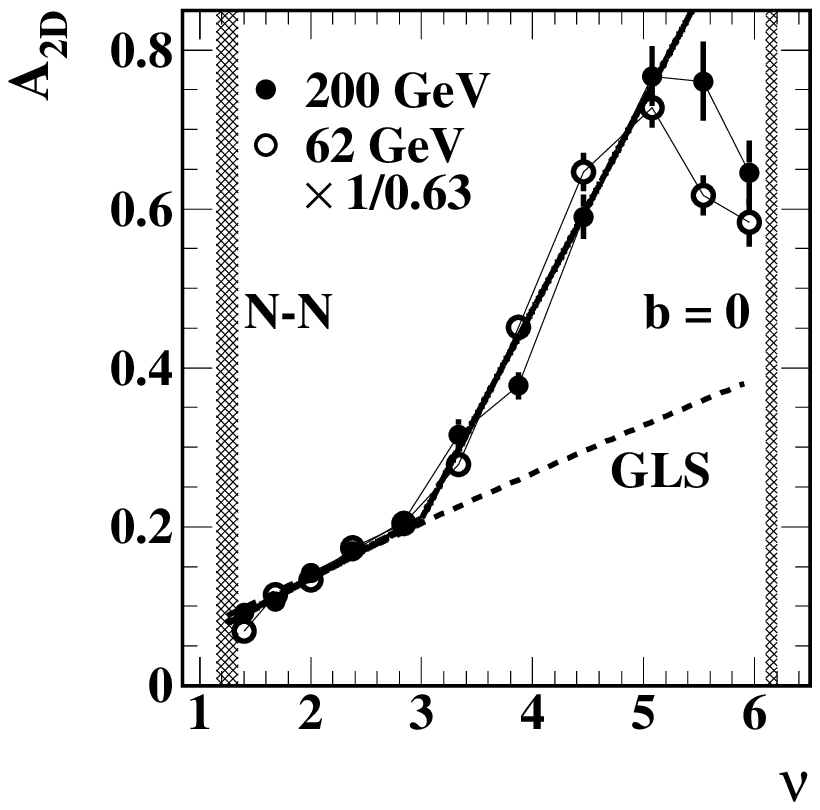}
  \includegraphics[width=.24\textwidth,height=.24\textwidth]{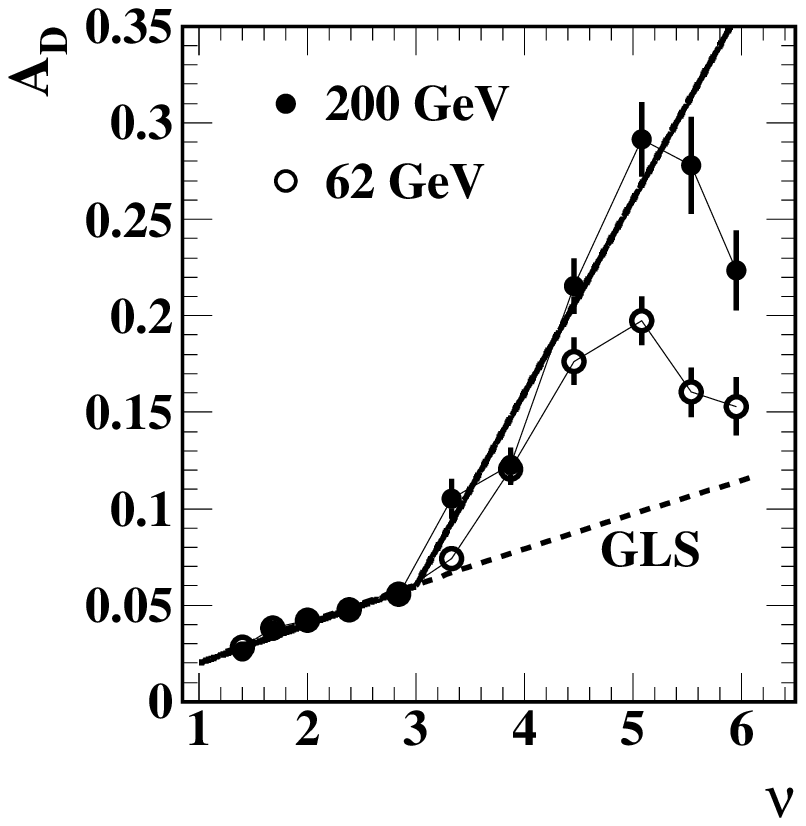}
}
\caption{
First: 200 GeV \auau correlations for 85-95\% central,
Second:  0-5\% central,
Third: Same-side 2D peak amplitude,
Fourth: Away-side 1D peak amplitude.
}
\label{aacorr}
\end{figure}

Figure~\ref{aacorr} (first, second) shows angular correlations for most-peripheral and most-central \auau collisions. The data extend down to 95\% central $\approx$ \nn collisions. The most-peripheral data correspond well with \pp results. Data from the peripheral region are required to establish a {\em Glauber linear superposition} (GLS) reference corresponding to {\em transparent} \aa collisions.

Figure~\ref{aacorr} (third, fourth) show results from 2D model fits that extract all information from the data~\cite{anomalous}. A simple parametrization includes three principal model elements: (a) SS 2D peak, (b) AS 1D peak on azimuth and (c) nonjet (NJ) azimuth quadrupole strongly evident in more-central \auau collisions -- a ``third component.'' The remaining structure (also modeled) is conversion electron pairs, Bose Einstein correlations and a 1D peak on $\eta_\Delta$ (soft component) that decreases to zero amplitude by mid centrality. The plots show the SS 2D peak and AS 1D peak amplitudes for two energies. 

The minimum-bias SS 2D peak is monolithic, with no separate ``ridge'' feature. The AS 1D peak on azimuth representing all aspects of transverse momentum conservation is dominated by parton scattering to back-to-back jets. The centrality dependence reveals a {\em sharp transition} in certain jet-related structure properties, a large change within one 10\% centrality bin in the slopes of SS and AS peak amplitudes (relative to GLS references) and the SS 2D peak $\eta$ width. The deviations from GLS for the SS 2D and AS 1D peaks are quantitative. Both remain consistent with jet-related structure.

A two-component study of \auau \pt spectra for identified hadrons was also conducted~\cite{hardspec}. The soft component maintains a fixed form (participant dissociation remains unchanged). Although the hard component changes substantially with centrality it is still described quantitatively by pQCD~\cite{fragevo}.

%\pt integral correlations (can also study \pt differential as in Ref.~\cite{davidhq2}).

%%%%%%%%%%%%%%%%%%5
\section{SS 2D peak and Fourier series and NJ quadrupole}

We now address the title issue: how to find or {\em not} find ``higher harmonic flows'' in \aa correlation data. In the past two years it has become popular to project all 2D angular correlations onto 1D azimuth, fit the projection with a Fourier series and interpret all series terms as ``harmonic flows.'' We now relate such procedures to the two- (or three-)component model.

\begin{figure}[htb]
\centerline{%
 \includegraphics[width=.24\textwidth,height=.24\textwidth]{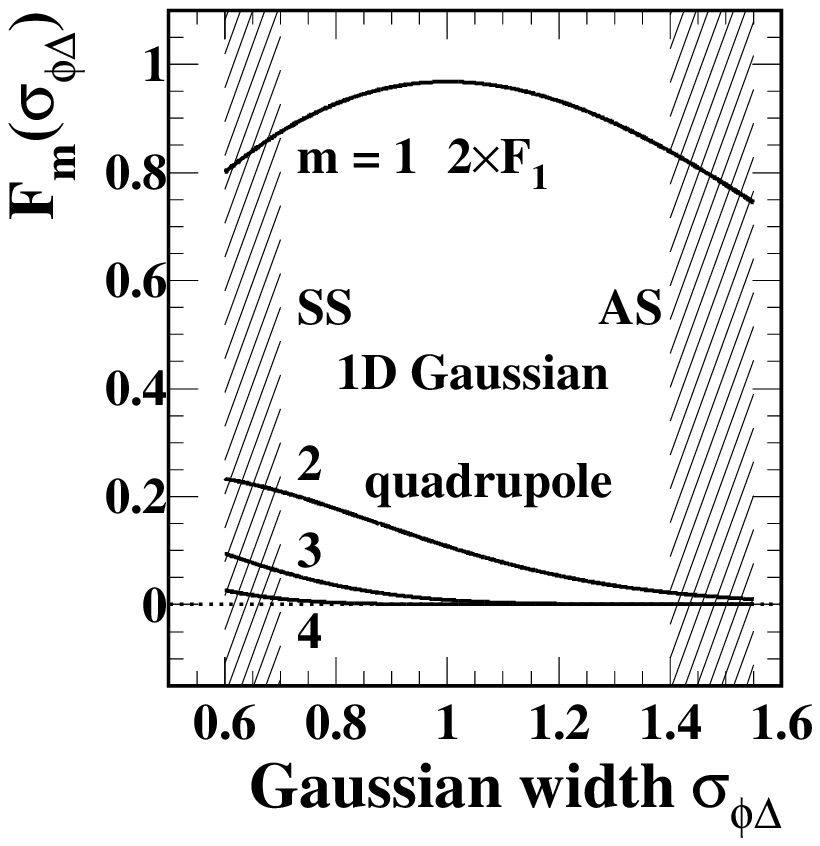} 
  \includegraphics[width=.24\textwidth,height=.24\textwidth]{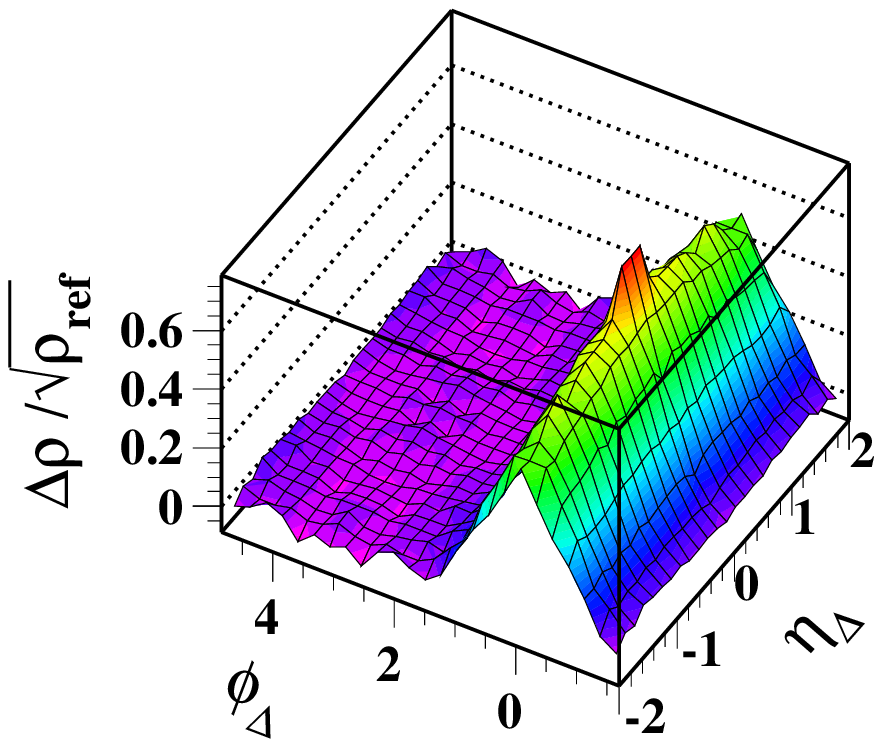} 
 \includegraphics[width=.24\textwidth,height=.24\textwidth]{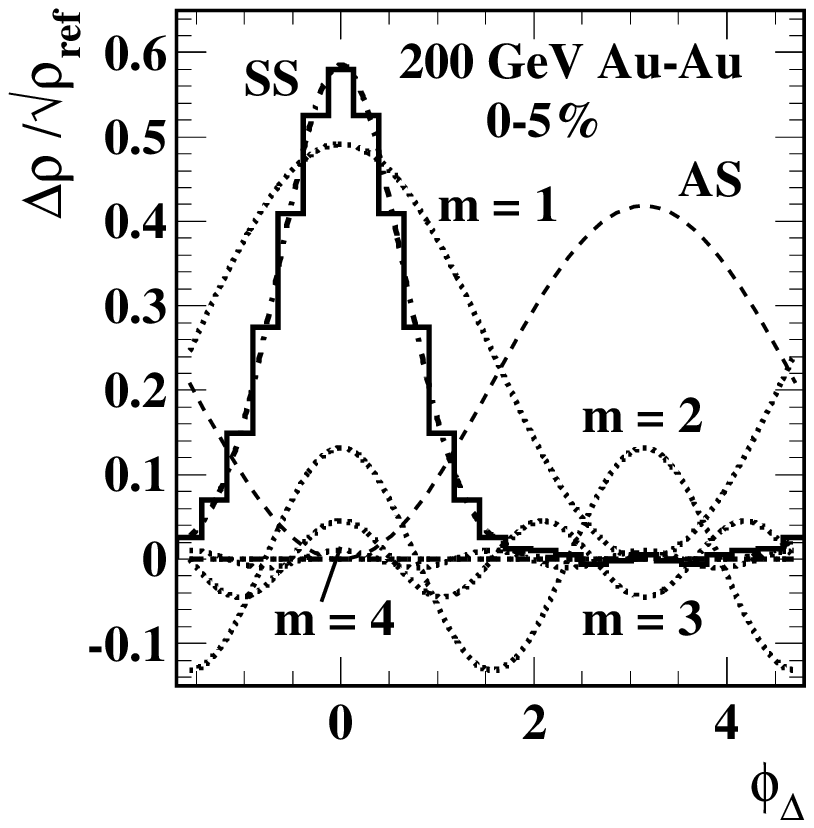}
} 
\caption{
First: Fourier coefficients vs width for 1D Gaussian,
Second: SS 2D peak for 0-5\% central \auau,
Third: SS peak 1D projection with Fourier components.
}
\label{fourier}
\end{figure}

Figure~\ref{fourier} (first) shows the Fourier coefficients for a Gaussian with r.m.s.\ width $\sigma_{\phi_\Delta}$ given by $F_m(\sigma_{\phi_\Delta}) = \sqrt{2/\pi} \,\sigma_{\phi_\Delta} \exp(-m^2 \sigma_{\phi_\Delta}^2 / 2)$~\cite{tzyam,multipoles}. The hatched bands labeled SS and AS correspond to the typical azimuth widths of SS and AS jet-related peaks. The AS 1D peak is described by a dipole ($m = 1$). The SS 2D peak requires several multipoles (Fourier series terms).

We next consider the azimuth structure of 2D correlations for 0-5\% central \auau collisions. For that centrality $A_Q\{2D\} \approx 0$~\cite{davidhq}. The 2D data histogram minus the fitted AS dipole leaves only the SS 2D peak, as shown in the second panel. The fit residuals are consistent with statistics. The {\em only source} of "higher harmonics" in those data is the SS 2D (jet) peak.

Figure~\ref{fourier} (third panel) shows the 1D projection of the SS 2D peak (bold histogram) and its Fourier components (bold dotted curves) with amplitudes described by $2A_X\{SS\}(b) = F_m[\sigma_{\phi_\Delta}(b)] G[\sigma_{\eta_\Delta}(b),\Delta \eta] A_{2D}(b) \equiv 2 \rho_0(b) v_m^2\{SS\}(b)$.  Factor $G$ represents projection of the SS 2D peak onto 1D azimuth~\cite{multipoles}. X represents various multipoles (e.g., dipole D, quadrupole Q, sextupole S, octupole O) with $2m$ poles for {\em cylindrical} multipoles, and SS denotes multipoles derived from the projected SS 2D peak.

\begin{figure}[htb]
\centerline{%
 \includegraphics[width=.24\textwidth,height=.24\textwidth]{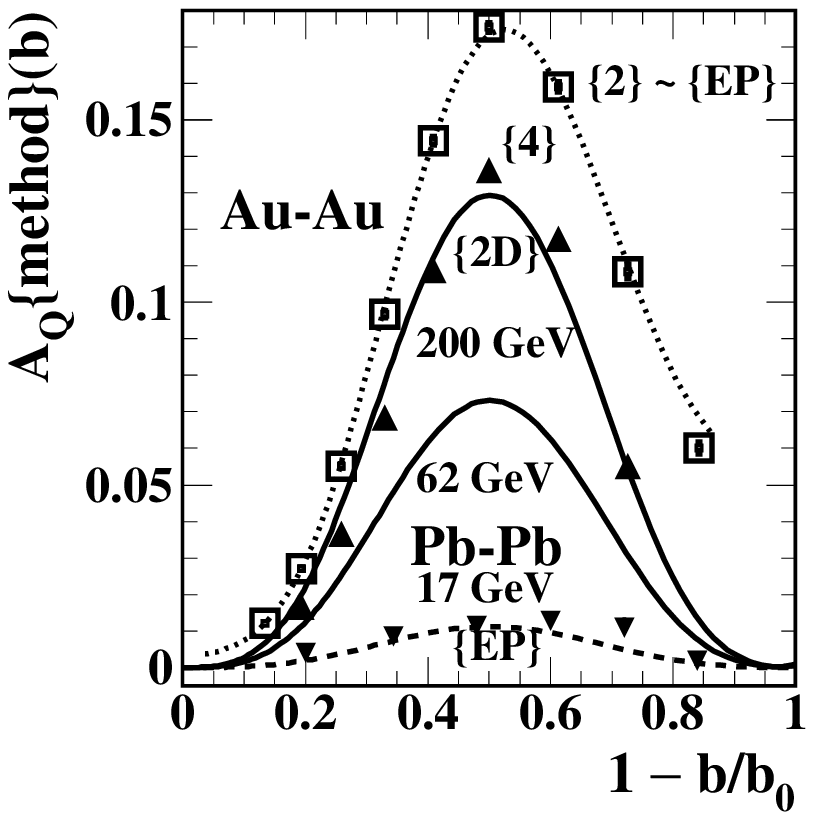}
  \includegraphics[width=.24\textwidth,height=.24\textwidth]{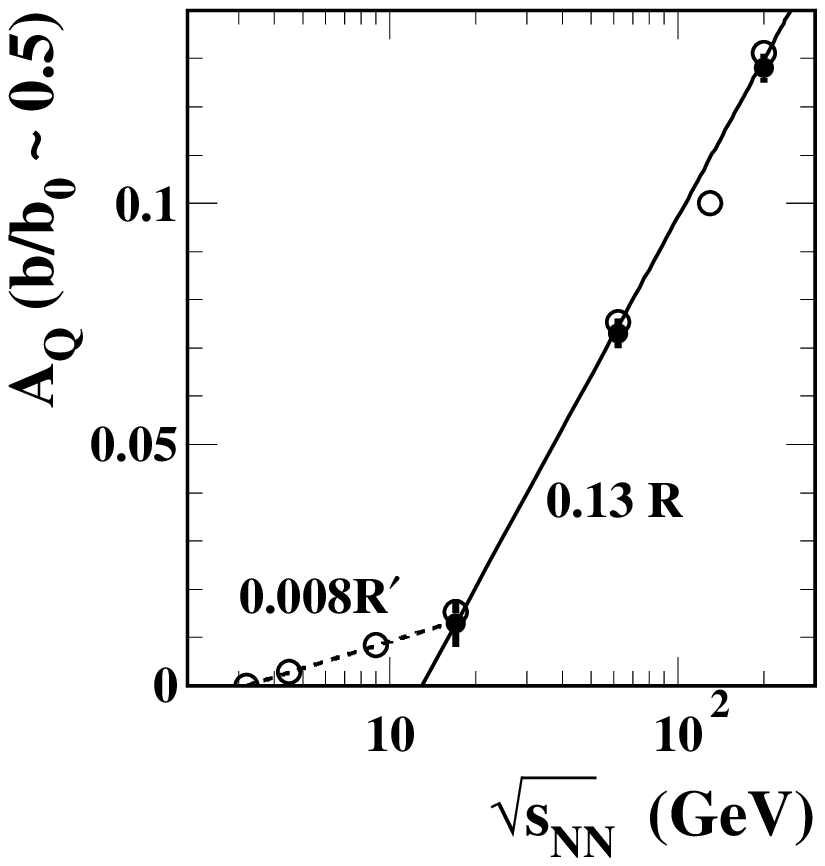}
 \includegraphics[width=.24\textwidth,height=.24\textwidth]{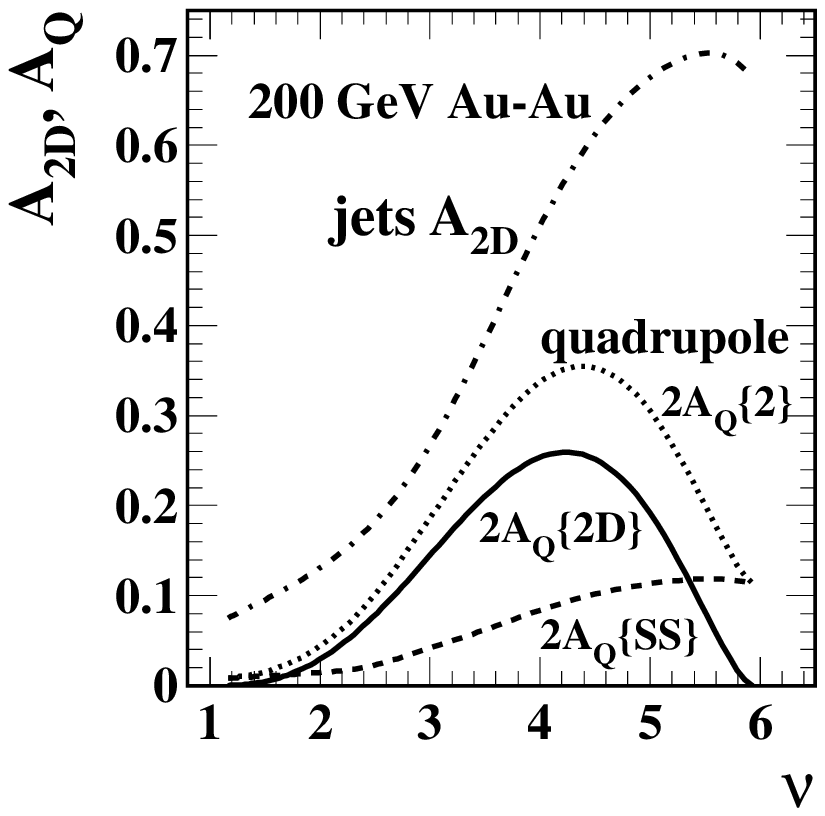}
}
\caption{
First: Quadrupole amplitudes for various $v_2$ methods,
Second: Quadrupole amplitude vs collision energy,
Third: Jet-related vs nonjet quadrupoles compared.
}
\label{njquad}
\end{figure}

Figure~\ref{njquad} (first panel) shows quadrupole amplitudes for various $v_2$ ``methods.'' The NJ quadrupole $A_Q\{2D\}$ is derived from 2D model fits with simple systematics described by
$A_Q\{2D\}(b) = 0.0045 N_{bin} R(\sqrt{s_{NN}}) \epsilon_{opt}^2 $ (solid and dashed curves)~\cite{davidhq}. $A_Q\{2D\}$ is related to the $v_m$ by $A_Q\{2D\}(b)\equiv \rho_0(b) v_2^2\{2D\}(b)$ with $\rho_0(b) \equiv n_{ch} / 2\pi \Delta \eta \approx  dn_{ch}/2\pi d\eta $. The NJ quadrupole is by definition distinct from SS and AS jet-related 2D correlation structure.

Figure~\ref{njquad} (second) shows energy variation proportional to $\log(\sqrt{s_{NN}})$ factors $R$, $R'$:  nucleon collectivity below 13 GeV and a novel (QCD?) phenomenon above 13 GeV carried by a small fraction of the final state~\cite{davidhq,quadspec}.

Figure~\ref{njquad} (third) shows the relation among jet-related structure, the NJ quadrupole and published $v_2$ data. The basic relation is $A_Q\{2\}(b) = A_Q\{2D\}(b) + A_Q\{SS\}(b)$. $A_Q\{SS\}(b)$ is derived entirely from fitted SS 2D peak properties, including strong width variations with centrality. We find that jet-related and nonjet quadrupole terms from a 2D model fit sum to published $v_2\{2\}$ data from the two-particle cumulant ``method''~\cite{2004}. The dotted curves in the first and third panels are the same, combining the measured NJ quadrupole~\cite{davidhq} and SS 2D jet peak~\cite{anomalous} trends. The \{2\} and \{EP\} methods are statistically equivalent~\cite{quadmeth}.  Note the small error bars within the open squares for $v_2\{2\}$ data in the first panel.  The published uncertainties have been multiplied by 5 to make them visible in the figure. Deviations from  $A_Q\{2D\}$ data (upper solid curve) are tens of error bars.

%1D Fourier series cannot describe the 2D data, is falsified by the data.

%\cite{davidhq,,multipoles}

%%%%%%%%%%%%%%
\section{Azimuth multipoles vs pQCD jets}

We now consider examples of finding ``higher harmonic flows.'' The recipe: (a) project all 2D angular correlations onto 1D azimuth, (b) perform a Fourier series fit to all data (which must describe {\em any} distribution on periodic azimuth), (c) interpret all Fourier series terms as ``flows.'' In effect, the SS 2D jet peak is ``fragmented'' (via Fourier series) to become "flows."

\begin{figure}[htb]
\centerline{%
 \includegraphics[width=.24\textwidth,height=.24\textwidth]{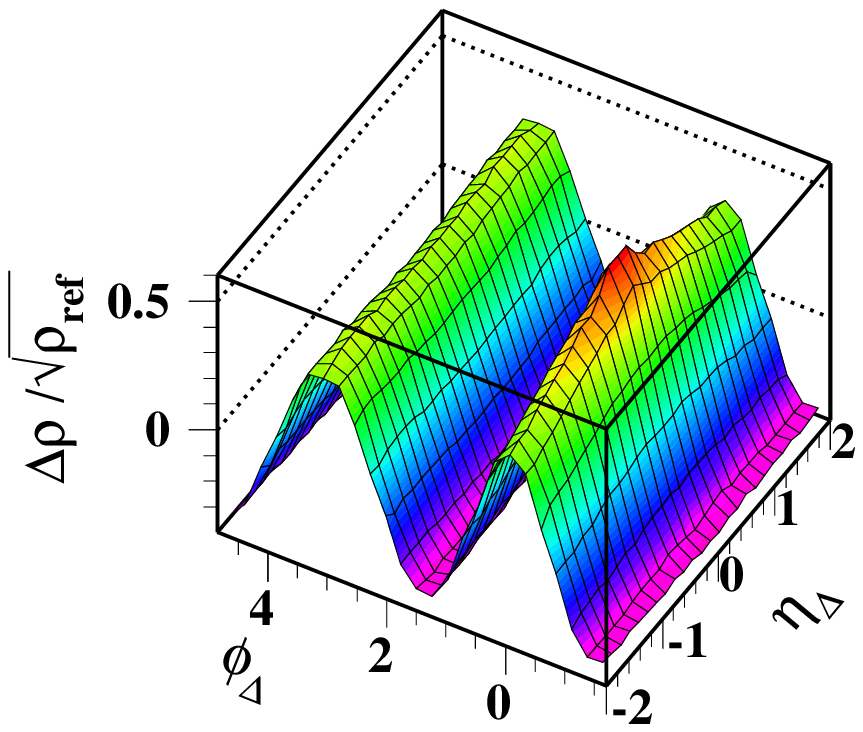}
 \includegraphics[width=.24\textwidth,height=.24\textwidth]{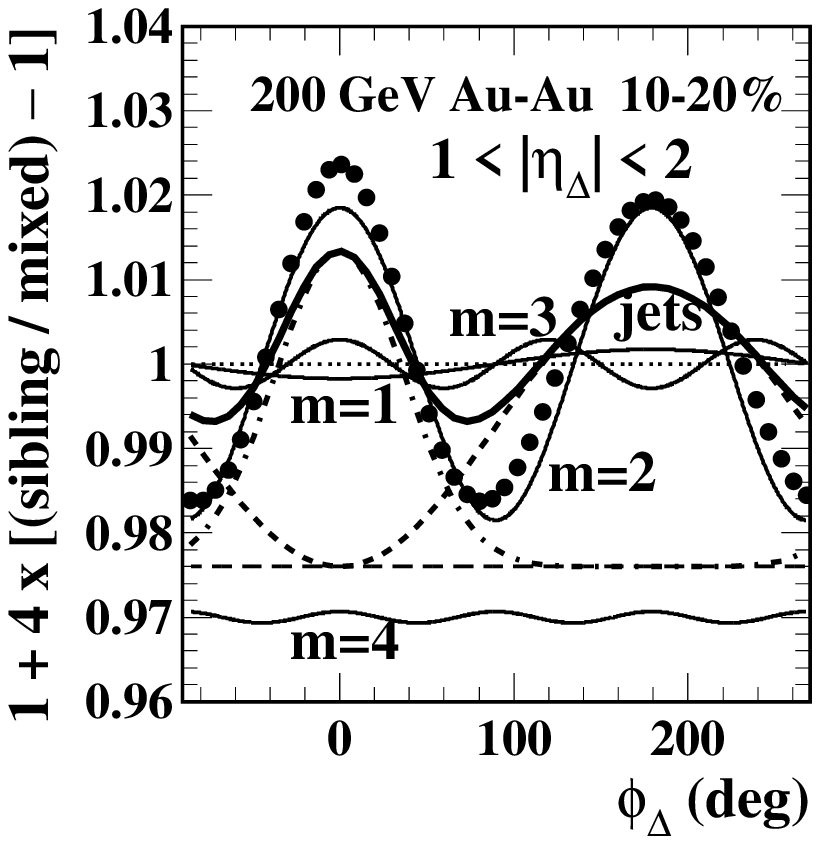}
 \includegraphics[width=.24\textwidth,height=.24\textwidth]{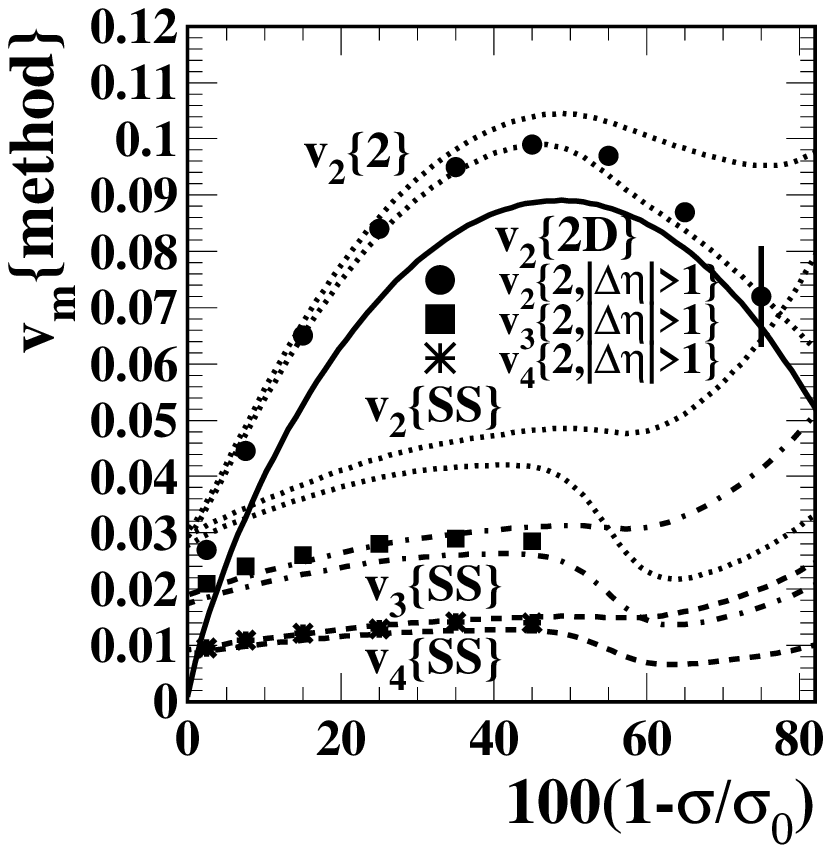}
}
\caption{
First:  Correlations for 10-20\% central Au-Au,
Second: Projection with Fourier coefficients,
Third: Comparing predicted multipoles with LHC data.
}
\label{multipoles}
\end{figure}

Figure~\ref{multipoles} (first, second) show 2D angular correlations from 10-20\% central \auau (2D histogram) projected onto 1D azimuth (solid points) and fitted with a four-term Fourier series (thin solid curves). The $m=3$ (sextupole) Fourier term is interpreted as ``triangular flow''\cite{gunther}. The sextupole structure (aka Mach Cones~\cite{tzyam}) is derived from the jet-related SS 2D peak.

Figure~\ref{multipoles} (third) shows calculated $v_m\{SS\}$ trends for SS 2D peak data from Ref.~\cite{anomalous} (broken curves) and the nonjet quadrupole trend $v_2\{2D\}$ from Ref.~\cite{davidhq} (solid curve).  The predicted $v_2\{2\}$ (upper dotted curves) accurately describes published data from Ref.~\cite{2004}. The points show a comparison of predictions with LHC data from Ref.~\cite{alice}. The agreement is excellent modulo an overall factor 1.3 increase~\cite{multipoles}. From the data we conclude that jet and nonjet-quadrupole trends are similar at RHIC and LHC energies.  Consequences of adding a sextupole term to 2D model fits or applying such analysis below the \auau sharp transition are further described in Ref.~\cite{sextupole}

\section{pQCD folding integral and fragment distributions}

We now demonstrate that the structure underlying claims of ``higher harmonic flows'' is described quantitatively by pQCD. To describe fragment yields in nuclear collisions requires a fragmentation function (FF) {\em ensemble} over the parton spectrum and the pQCD parton spectrum itself. The FF ensemble on dijet energy $Q$ is accurately described (to statistical limits) by a {\em beta} distribution with rescaling of the total rapidity~\cite{ffprd}. FFs are represented by $D(x,Q^2) \rightarrow D(y,y_{max}) = 2 n_{ch}(y_{max})\beta[u;p(y_{max}),q(y_{max})]$. Model parameters $(p,q)$ are nearly constant over a large energy range. The dijet fragment multiplicity $2n_{ch}(y_{max})$ is simply determined by energy conservation and scales with jet energy approximately as $y_{max}^2 \equiv \ln^2(Q/m_\pi)$.

\begin{figure}[htb]
\centerline{%
\includegraphics[width=.24\textwidth,height=.24\textwidth]{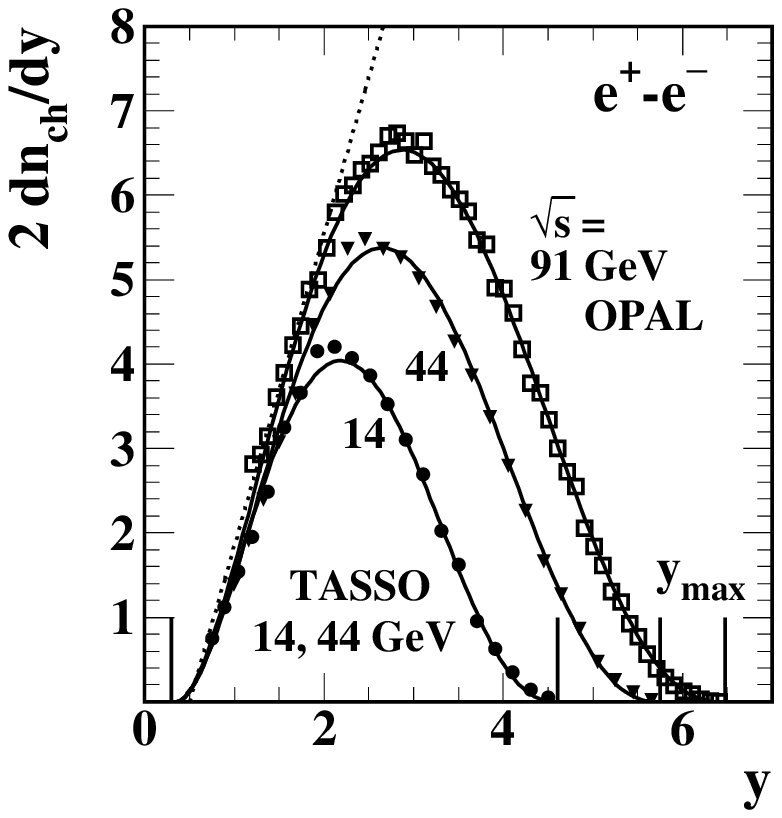}
\includegraphics[width=.24\textwidth,height=.24\textwidth]{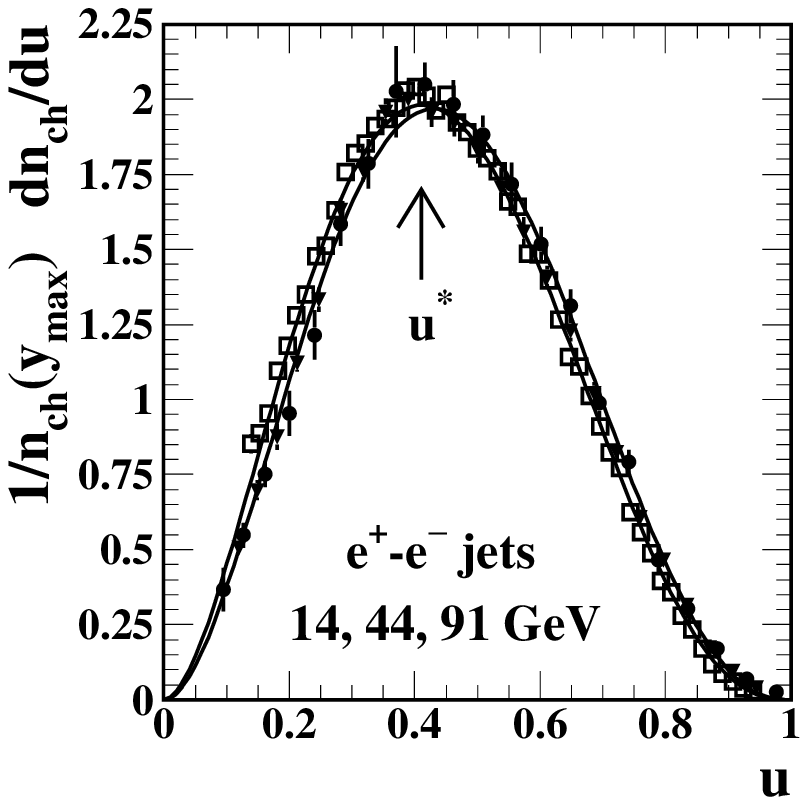}
\includegraphics[width=.24\textwidth,height=.24\textwidth]{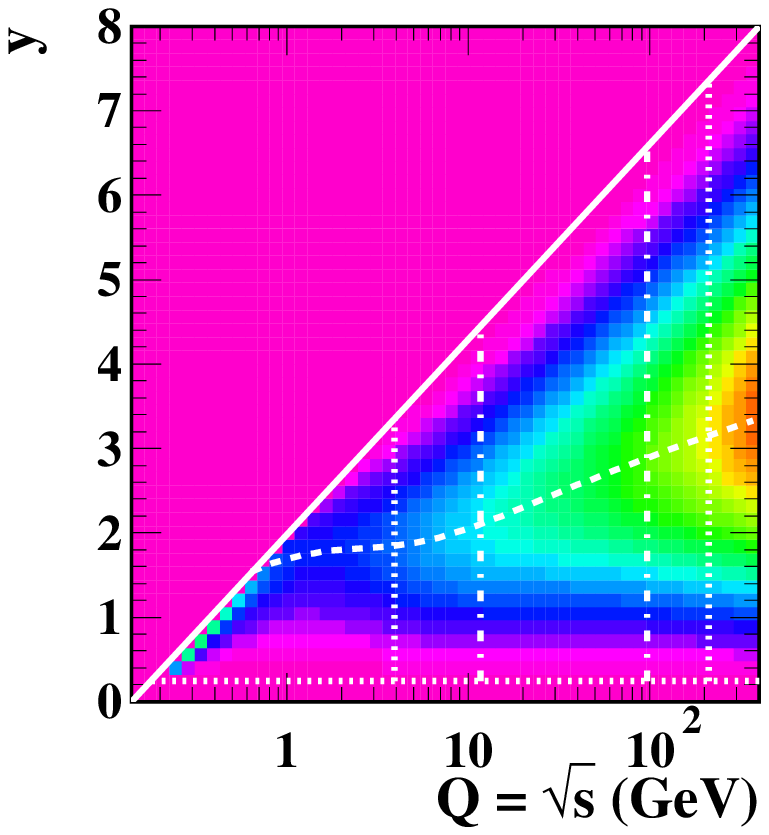}
\includegraphics[width=.24\textwidth,height=.237\textwidth]{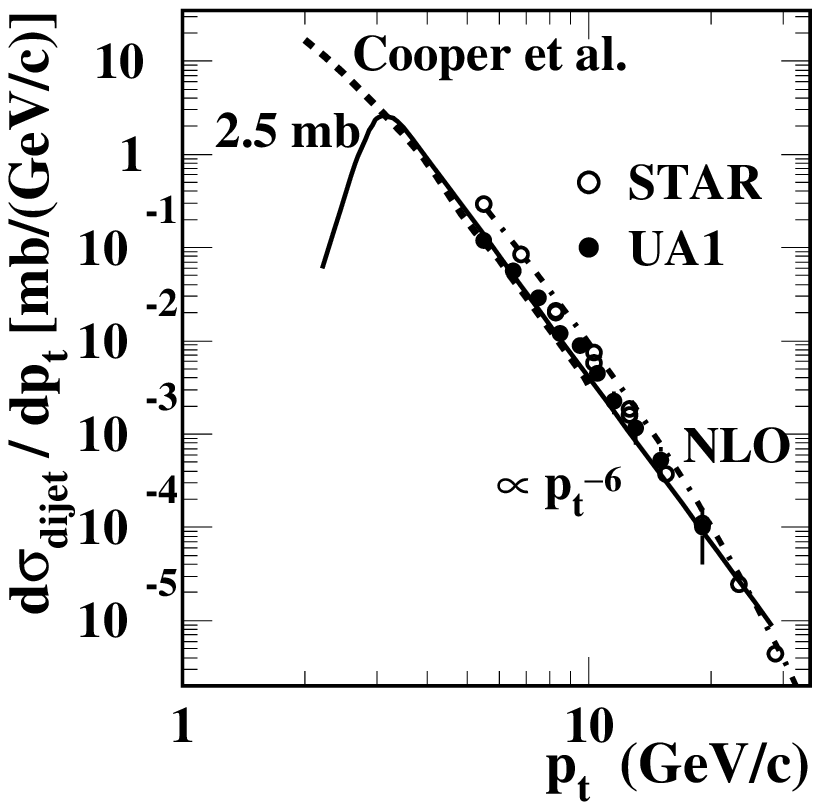}
}
\caption{
First: Measured fragmentation functions (FFs),
Second: FFs rescaled and beta distribution,
Third: FF ensemble vs energy $Q$,
Fourth: pQCD parton spectra.
}
\label{pqcd}
\end{figure}

Figure~\ref{pqcd} (first, second) show measured FFs on rapidity $y\equiv\ln[(p+E)/m_\pi]$ for three dijet energies and the same data rescaled on normalized rapidity $u \approx y/y_{max}$~\cite{ffprd}. For jet energies relevant to nuclear collisions more than half the jet fragments fall below $p_t = 1$ GeV/c ($y \approx 2.7$), falsifying the assumption that hydro dominates below 2 GeV/c. The third panel shows the parametrized ensemble of FFs over a large energy range up to LEP II.

Figure~\ref{pqcd} (fourth) shows a parametrized parton spectrum (solid curve) compared to jet spectrum data (points) and pQCD theory calculations (other curves). pQCD describes the parton spectrum down to Q = 6 GeV (3 GeV jets) corresponding to perturbative length scale 0.03 fm. The parton spectrum folded with FFs as the {\em measured} non-perturbative aspect of QCD determine {\em fragment distributions} (FDs) down to zero hadron momentum~\cite{fragevo}. pQCD FDs can then be compared with spectrum hard components $H$~\cite{hardspec}.

%DGLAP describes certain aspects of FFs at larger momentum, but that is irrelevant for this problem.

%Those pQCD tools can be used to describe spectrum hard components accurately. First for \pp collisions.

\section{pQCD descriptions of spectrum hard components}

\begin{figure}[htb]
\centerline{%
 \includegraphics[width=.24\textwidth,height=.237\textwidth]{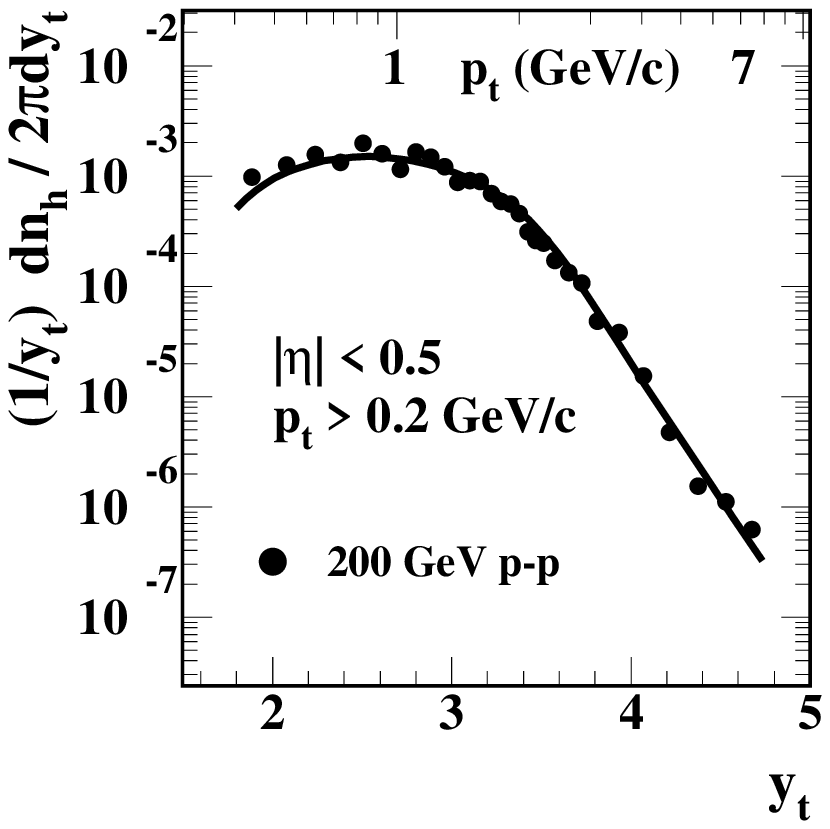}
  \includegraphics[width=.24\textwidth,height=.24\textwidth]{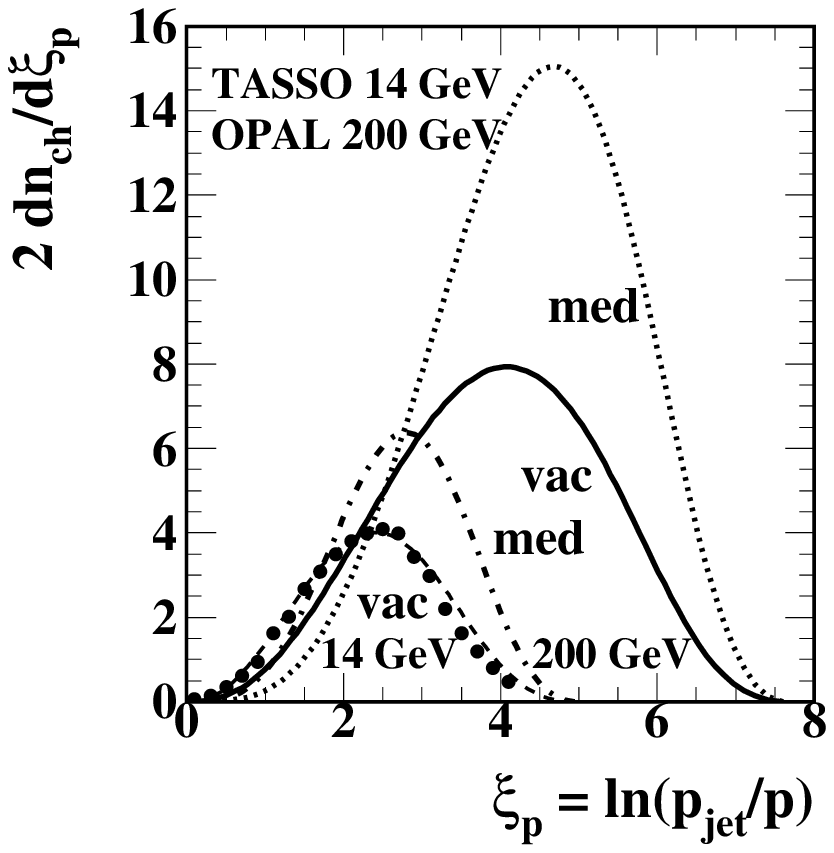} 
 \includegraphics[width=.24\textwidth,height=.24\textwidth]{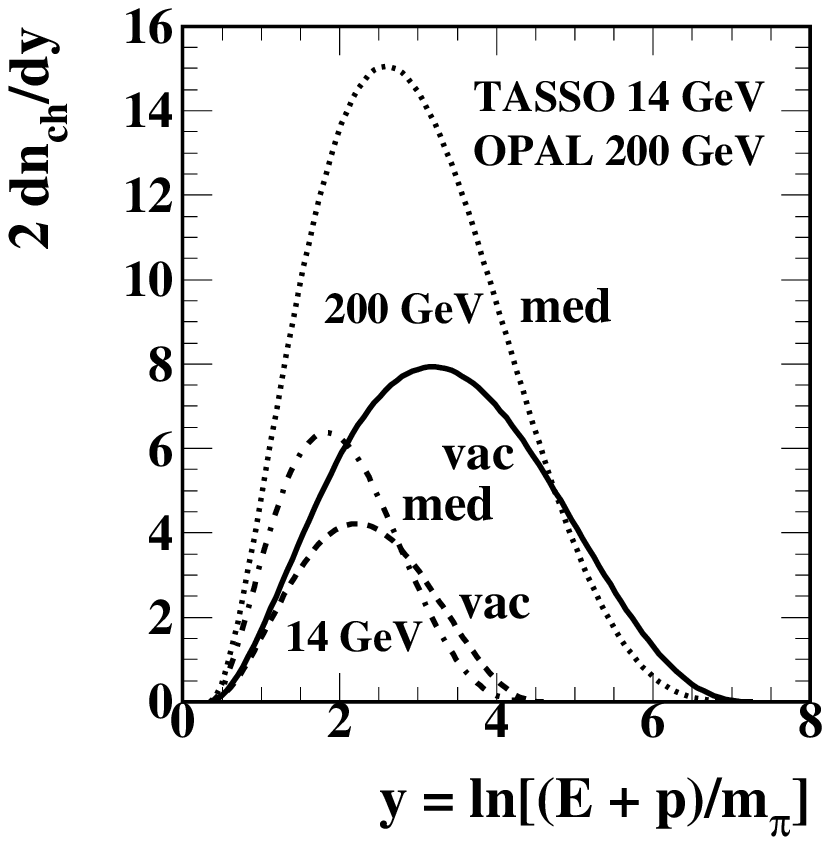}
  \includegraphics[width=.24\textwidth,height=.236\textwidth]{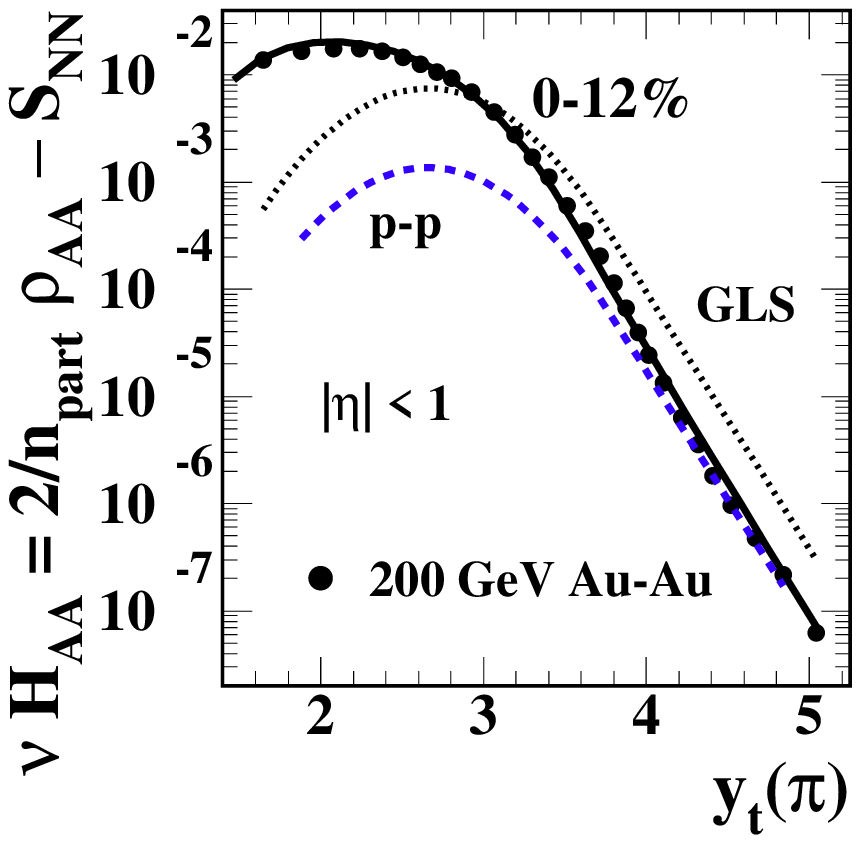} 
}
\caption{
First: pQCD calculated fragment distribution (FD) vs \pp data,
Second: Modified FFs  on $\xi_p$,
Third: Modified FFs on $y$,
Fourth: pQCD FD vs \auau data
}
\label{fddist}
\end{figure}

Figure~\ref{fddist} (first) shows a calculated FD (solid curve,~\cite{fragevo}) compared to spectrum hard component $H$ derived from 200 GeV non-single-diffractive \pp data~\cite{ppprd}. A single parameter in the pQCD spectrum, the lower bound on the spectrum near 3 GeV, has been adjusted to accommodate the data. The data description is excellent. Note the mode of the FD near 1 GeV/c.

In \pp collisions the two-component spectrum model is $(1/n_s) dn_{ch}/y_t dy_t = S_{0}(y_t) + (n_h/n_s) H_0(y_t)$~\cite{ppprd}. In \aa collisions the model is generalized to $(2/N_{part}) dn_{ch}/y_t dy_t = S_{NN}(y_t) + \nu H(y_t,\nu)$,
with $n_h/n_s \leftrightarrow \nu = 2N_{bin}/N_{part}$ as analogous ``centrality'' parameters~\cite{hardspec}. We expect hard component $H(y_t,\nu)$ to be modified in more-central \aa collisions due to modification of FFs. FF shape modification is modeled by changing parameter $q$ in $\beta(u;p,q)$ by $\Delta q$. Fragment number $2n_{ch}$ is rescaled so as to conserve the parton energy.

Figure~\ref{fddist} (second, third) illustrates FFs for two dijet energies without (vacuum) and with (medium) modification. Figure~\ref{fddist} (fourth) shows the measured spectrum hard component for pions from 0-12\% central 200 GeV \auau collisions (points, \cite{hardspec}) and the calculated FD with modified FFs (solid curve,~\cite{fragevo}). Parameter $\Delta q$ ($\approx 1$) has been adjusted to describe the five-fold suppression at $y_t = 5$ ($p_t \approx 10$ GeV/c). All else remains the same as for the first panel. The description of \auau data is remarkable. The dotted curve labeled GLS is the FD prediction for no FF modification.

\section{Resolved minijets and hadron yields vs \auau centrality}

\begin{figure}[htb]
\centerline{%
 \includegraphics[width=.24\textwidth,height=.24\textwidth]{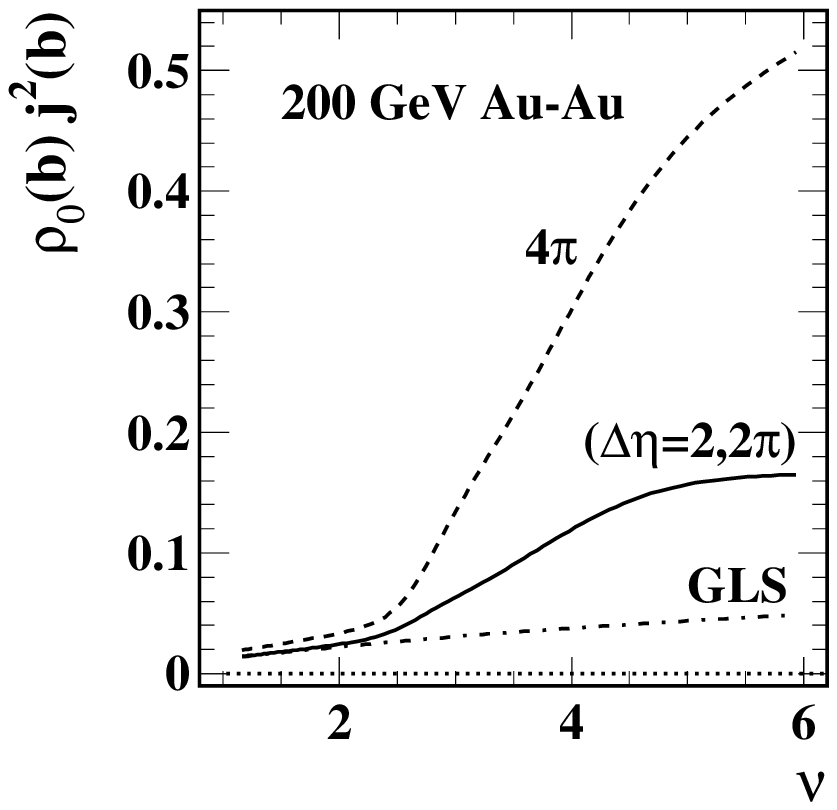}
 \includegraphics[width=.24\textwidth,height=.24\textwidth]{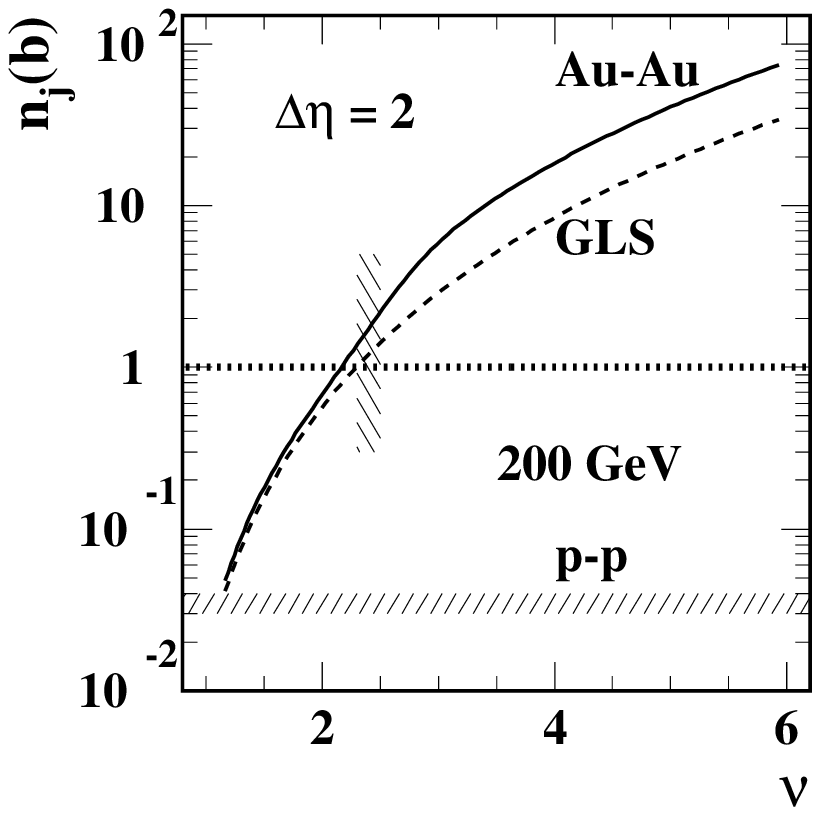}
 \includegraphics[width=.24\textwidth,height=.24\textwidth]{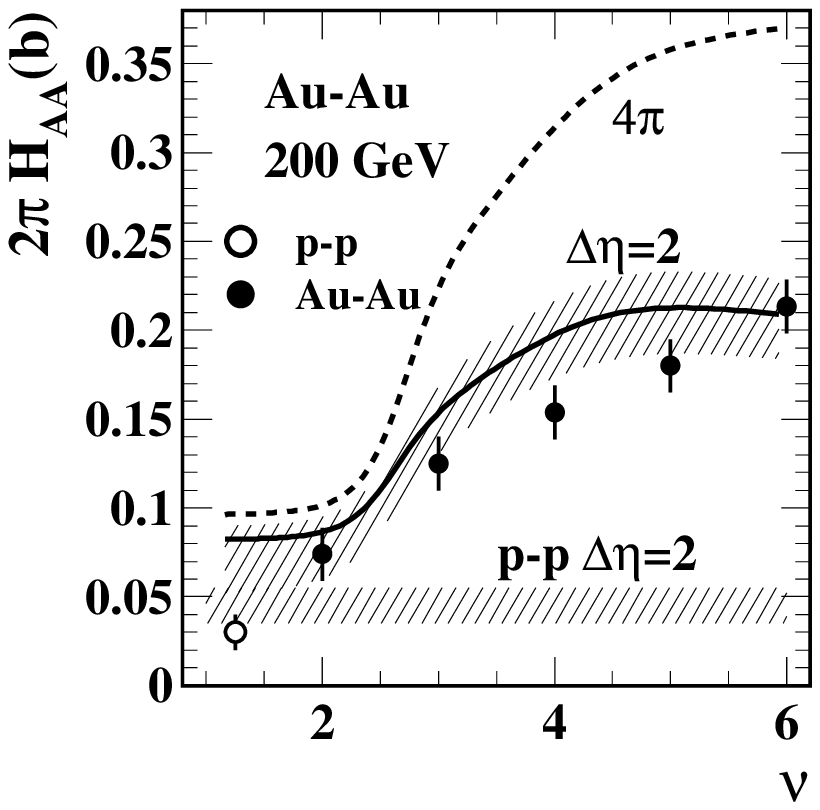}
  \includegraphics[width=.24\textwidth,height=.24\textwidth]{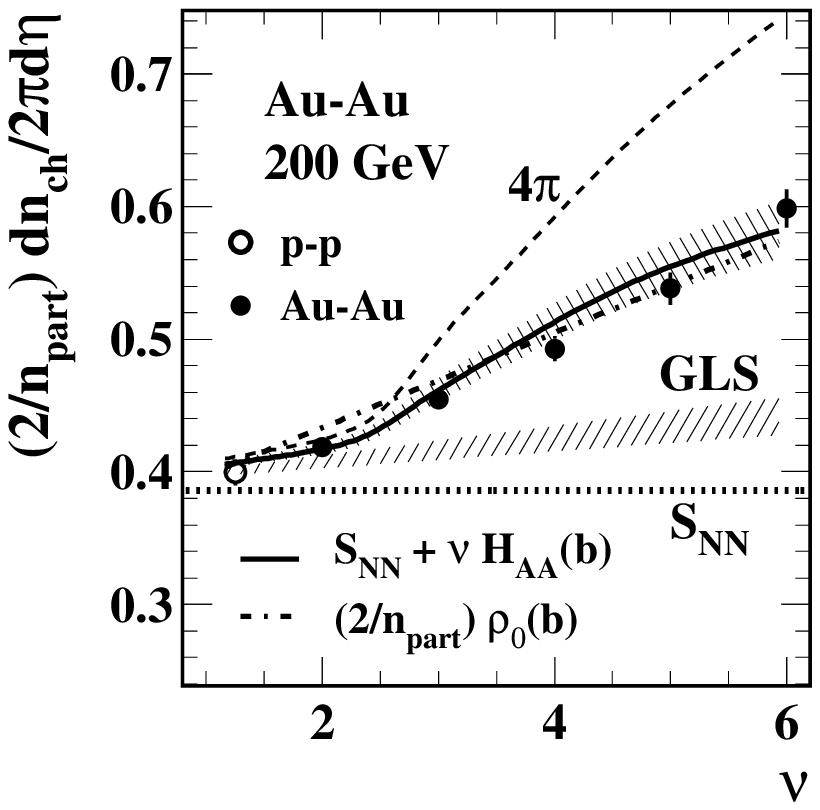}
}
\caption{
First: SS 2D peak volume,
Second: pQCD calculated dijet frequency
Third:  Fragment yield vs \auau centrality
Fourth: Total hadron yield vs centrality.
}
\label{pqcdhard}
\end{figure}

Figure~\ref{pqcdhard} demonstrates the correspondence between jet-related angular correlations and hard-component hadron yields with \auau centrality. The first panel shows the SS 2D peak volume within acceptance $\Delta \eta = 2$ (solid curve). The second panel shows the pQCD predicted dijet frequency within $\Delta \eta$. By combining the two trends the mean jet fragment multiplicity can be inferred. Recombining the fragment multiplicity with the dijet frequency predicts the hard component yields (solid curve) in the third panel and (combined with fixed soft component $S_{NN}$) the total hadron yields in the fourth panel. The prediction (solid curve, \cite{jetspec}) is compared with the measured hadron yields from Ref.~\cite{hardspec} (points). Again the agreement is remarkable.

% and demonstrates that one third of the hadronic final state is contained within resolved jets.

\section{Summary}

This presentation reviews quantitative relations among (a) jet-related 2D angular correlations, (b) jet-related spectrum hard components and (c) pQCD-calculated jet frequencies and fragment distributions that strongly support a jet interpretation of the SS 2D peak for all \auau centralities. A pQCD description of hadron production at RHIC agrees with spectrum data within their uncertainties  and implies that one third of final-state hadrons in central \auau collisions are contained {\em within resolved jets}.
The more-peripheral 50\% of the \auau total cross section corresponds to {\em transparent}  collisions where dijet characteristics are just as for \pp collisions. Over the same transparency interval the NJ azimuth  quadrupole measured by a statistically equivalent quantity increases to 2/3 of its maximum value, making a conventional hydro interpretation very unlikely.
Proponents of ``higher harmonic flows'' must confront the many results that falsify a flow hypothesis and support a jet mechanism for the dominant correlation and spectrum structure. Cherry picking data features and favoring analysis methods and ``theories'' that seem to support a preferred (flow) hypothesis while disregarding contrasting results from pQCD is questionable practice.

%%%%%%%%%%%%%%%%%%

\end{document}